\documentclass{article}
\usepackage{spconf,amsmath,graphicx,hyperref}
\usepackage{booktabs}
\usepackage{amssymb}
\usepackage{braket}
\usepackage{tabularx}

\title{Batched Training for QLSTM vs. QFWP: A System-Oriented Approach to EPC-Aware RMSE-DA}

\name{
Jun-Hao Chen$^{1}$, Ming-Kai Hung$^{1}$, Yun-Cheng Tsai$^{1}$, Samuel Yen-Chi Chen$^{2}$%
}
\address{
$^{1}$Department of Technology Application and Human Resource Development,\\
National Taiwan Normal University, Taipei, Taiwan\\
$^{2}$Wells Fargo Bank, USA
}

\begin{document}
%\ninept
%
\maketitle
\begin{abstract}
We compare two quantum sequence models, QLSTM and QFWP, under an Equal-Parameter-Count (EPC; trainable-parameter) and adjoint-differentiation setup on daily EUR/USD forecasting as a controlled 1D time-series case study. Across 10 random seeds and batch sizes $4$--$64$, we measure component-wise runtimes (train-forward, backward, full-train, and inference) and accuracy (RMSE and Directional Accuracy (DA)). Batched forward scales well ($\sim 2.2$--$2.4\times$), but backward scales modestly (QLSTM $\sim 1.01$--$1.05\times$, QFWP $\sim 1.18$--$1.22\times$), capping end-to-end training speedups near $2\times$. QFWP attains lower RMSE and higher DA at all batches (Wilcoxon $p \le 0.004$, large Cliff's $\delta$), while QLSTM reaches the highest throughput at batch $64$, revealing a clear speed--accuracy Pareto frontier. We provide an EPC-aligned, numerically checked benchmarking pipeline and practical guidance on batch choices; broader datasets and hardware/noise settings are left for future work.
\end{abstract}

\section{Introduction}\label{sec:introduction}
Quantum sequence models have gained traction for signal processing tasks that require learning long-range dependencies~\cite{schuld2015introduction,Schuld2019,parcollet2019bidirectional}. Among these, QLSTM augments classical memory with variational quantum circuits~\cite{chen2022quantum}, while programmable architectures such as QFWP avoid recurrent quantum states by using a slow classical network to configure shallow circuits at each step~\cite{chen2024learning,liu2025programming}. Despite this progress, systematic benchmarking under batching and adjoint differentiation remains limited. Prior work emphasizes architectural efficiency~\cite{dehghani2021efficiency,chen2022quantum_qcnn}, yet few studies quantify forward and backward scaling or accuracy stability across batch sizes~\cite{hoch2024variational}. 

This paper closes these gaps by presenting a unified Equal-Parameter-Count (EPC) benchmark of QLSTM and QFWP on a canonical OHLC$\rightarrow$next-day-close forecasting task, with EUR/USD as a representative 1D case study for controlled system measurement. We define EPC over \emph{trainable} parameters, report the quantum/classical split, and decompose wall-clock time into train-forward, backward, full-train, and inference components to expose adjoint-gradient bottlenecks. Forward pass speedups scale strongly with batching, while backward pass scaling remains modest, yielding end-to-end gains of $\sim 2\times$. Under this fixed-epoch training budget, QFWP achieves lower RMSE and higher DA, whereas QLSTM provides the highest throughput, forming a clear speed--accuracy Pareto frontier.

\section{Methodology}
\label{sec:methodology}

\textbf{Task and Data.}
We use daily EUR/USD OHLC (Dec 2003--Dec 2024, seq-len 5) with an 80/20 train--test split.
Each rolling window is min--max normalized; the same scale is applied to the target.
We report RMSE and Directional Accuracy (DA) on the original scale:
\[
\mathrm{DA}=\tfrac{100}{|\mathcal{I}|}\!\sum_{t\in\mathcal{I}}\!\mathbf{1}[\mathrm{sign}(\hat{y}_t-\hat{y}_{t-1})=\mathrm{sign}(y_t-y_{t-1})],
\]
where $\mathcal{I}=\{t:y_t\neq y_{t-1}\}$.
We use a single 1D dataset to isolate system effects.

\textbf{Quantum Models.}
QLSTM~\cite{chen2022quantum} replaces each LSTM~\cite{hochreiter1997long} gate with a shallow VQC~\cite{cerezo2021variational} operating on $[\mathbf{x}_t;\mathbf{h}_{t-1}]$, where $\mathbf{x}_t\!\in\!\mathbb{R}^4$ and $\mathbf{h}_{t-1}\!\in\!\mathbb{R}^H$.
Each gate outputs $\mathbf{e}_t^{(g)}=\mathrm{VQC}_g([\mathbf{x}_t;\mathbf{h}_{t-1}];\Theta_g)$ and uses $\sigma/\tanh$ to update
$\mathbf{c}_t=\mathbf{f}_t\odot\mathbf{c}_{t-1}+\mathbf{i}_t\odot\tilde{\mathbf{c}}_t$ and
$\mathbf{h}_t=\mathbf{o}_t\odot\tanh(\mathbf{c}_t)$, with $y_t=W_{\mathrm{out}}\mathbf{h}_t+b_{\mathrm{out}}$.
Each VQC (depth 1) initializes $n_q=4+H$ wires in $\ket{+}$, encodes via $R_y$, applies an even--odd CNOT ladder, and measures $\mathbb{E}[Z]$ on the first $H$ wires.
With $H=3$, QLSTM has 32 trainable parameters (28 quantum + 4 classical).

QFWP~\cite{chen2024learning} follows fast-weights~\cite{schmidhuber1992learning,ba2016using}: a slow classical network \emph{programs} a shallow circuit each step.
Given $\mathbf{s}_t$ (OHLC), it computes $\mathbf{z}_t=W_e\mathbf{s}_t+b_e$, selectors
$\boldsymbol{\alpha}_t=W_\ell\mathbf{z}_t+b_\ell$ and $\boldsymbol{\beta}_t=W_q\mathbf{z}_t+b_q$, and updates angles
$\Theta_t=\Theta_{t-1}+\boldsymbol{\alpha}_t\boldsymbol{\beta}_t^\top$ ($D=1$, $Q=3$).
The VQC encodes $\mathbf{s}_t$ with $R_y$, applies one entangling layer, measures $A=1$, and predicts $\hat{y}_t=W_p\mathbf{a}_t+b_p$.
All 33 trainable parameters are classical; per-step angles are generated by the slow net (see Table~\ref{tab:epc_detailed}).
Both models are EPC-aligned and implemented in PennyLane~\cite{bergholm2018pennylane} (\texttt{lightning.qubit}) with PyTorch~\cite{paszke2019pytorch} autograd (\texttt{diff\_method=adjoint}).

\textbf{Training and Evaluation Protocol.}
We compare non-batch vs batch sizes $\{4,8,16,32,64\}$ over 10 seeds using identical splits and synchronized initialization.
For each run we time: (1) Train-Forward, (2) Backward, (3) Full-Training (2 epochs, fixed budget), and (4) Inference-Forward (after warm-up).
Speedup is $T_{\text{non-batch}}/T_{\text{batch}}$ summarized as median [IQR]; accuracy is mean$\pm$std.
We fix epochs (not steps) to emphasize throughput, yielding fewer updates at larger batches (4,374 samples $\rightarrow$ 1,094/547/274/137/69 steps per epoch for batch 4/8/16/32/64). No gradient accumulation or LR scaling is used.

\textbf{Numerical Checks and Statistics.}
We feed identical tensors to batched/non-batched implementations.
For QLSTM we verify forward equivalence ($L_2\le 10^{-6}$) and measure backward wall-time; for QFWP we verify forward equivalence only (per-sample adjoint cost).
Across seeds, paired Wilcoxon signed-rank tests~\cite{wilcoxon1945individual} compare QLSTM vs QFWP per batch, with Cliff's $\delta$~\cite{cliff1993dominance} effect sizes; timing medians [IQR] reveal scaling bottlenecks.

All experiments use a state-vector simulator (no hardware execution or noise modeling); timings therefore reflect relative simulation efficiency.

\section{Experiments and Results}
\label{sec:exp_and_res}

% New figure: 更新時間 2025-10-15 11:00
\begin{figure}[t]
\centering
\includegraphics[width=\linewidth]{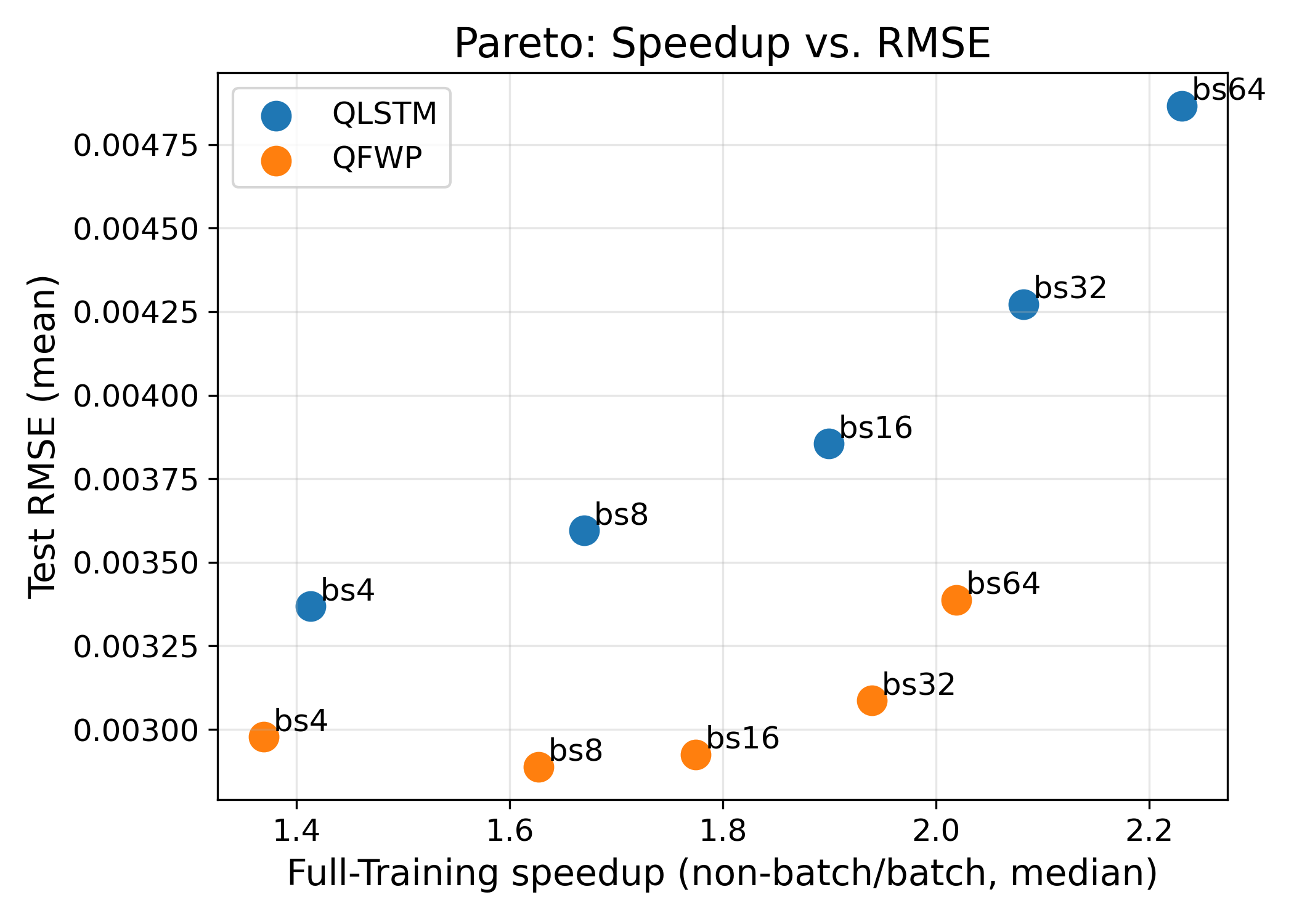}
\caption{Pareto view: full-training speedup (median) vs.\ RMSE (mean). Each point is annotated with its batch size, and marker styles distinguish QLSTM vs.\ QFWP for grayscale readability.}
\label{fig:pareto_frontier}
\end{figure}

% --- EPC Alignment mini-table (drop-in, compiles as-is) ---
\begin{table*}[!tbhp]
  \centering
  \small
  \caption{Detailed parameter breakdown for EPC alignment.}
  \label{tab:epc_detailed}
  \begin{tabular*}{\textwidth}{@{\extracolsep{\fill}}lrrrr}
    \toprule
    Model & Total & Quantum & Classical & Architecture \\
    \midrule
    \multicolumn{5}{l}{\textbf{QLSTM} (input=4, hidden=3, $q_\text{depth}=1$, $n_\text{qubits}=7$)} \\
    \quad 4 VQC gates & 28 & 28 & 0 & $4 \times (1 \times 7)$ \\
    \quad output\_post\_processing & 4 & 0 & 4 & Linear(3$\to$1) \\
    \quad \textit{Subtotal} & \textit{32} & \textit{28} & \textit{4} & --- \\
    \midrule
    \multicolumn{5}{l}{\textbf{QFWP} ($s_\text{dim}=4$, $a_\text{dim}=1$, latent=3, $n_\text{qubits}=3$, $q_\text{depth}=1$)} \\
    \quad slow\_program\_encoder & 15 & 0 & 15 & Linear(4$\to$3) \\
    \quad slow\_program\_layer\_idx & 4 & 0 & 4 & Linear(3$\to$1) \\
    \quad slow\_program\_qubit\_idx & 12 & 0 & 12 & Linear(3$\to$3) \\
    \quad post\_processing & 2 & 0 & 2 & Linear(1$\to$1) \\
    \quad \textit{Subtotal} & \textit{33} & \textit{0} & \textit{33} & --- \\
    \midrule
    \textbf{Difference (QLSTM - QFWP)} & \textbf{-1} & --- & --- & \textbf{3.03\%} \\
    \bottomrule
  \end{tabular*}
  \vspace{1mm}

  \footnotesize
  Note: QFWP's quantum circuit parameters (3 per sample per step: 1 depth $\times$ 3 qubits) are not trainable; they are computed dynamically by the slow-programmer network.
\end{table*}

% --- (A) Accuracy table: RMSE/DA with Wilcoxon p-values and Cliff's delta ---
% New table: 更新時間 2025-10-15 11:00
\begin{table*}[htbp]
  \centering
  \small
  \caption{Accuracy across batch sizes ($mean \pm std$ over seeds). Lower RMSE and higher DA are better. Per-batch significance uses Wilcoxon signed-rank test (QLSTM vs. QFWP); effect size reported as Cliff's $\delta$.}
  \label{tab:acc_main}
  \begin{tabular*}{\textwidth}{@{\extracolsep{\fill}}lrrrrrr}
    \toprule
    Batch & RMSE (QLSTM) & RMSE (QFWP) & $p$-value & $\delta$(RMSE) & DA\,(\%) (QLSTM) & DA\,(\%) (QFWP) \\
    \midrule
    4 & 0.0034 $\pm$ 0.0001 & \textbf{0.0030 $\pm$ 0.0002} & 0.004 & 0.800 & 69.04 $\pm$ 0.89 & \textbf{72.72 $\pm$ 1.88} \\
    8 & 0.0036 $\pm$ 0.0001 & \textbf{0.0029 $\pm$ 0.0001} & 0.002 & 1.000 & 67.89 $\pm$ 0.64 & \textbf{72.75 $\pm$ 0.93} \\
    16 & 0.0039 $\pm$ 0.0001 & \textbf{0.0029 $\pm$ 0.0001} & 0.002 & 1.000 & 65.56 $\pm$ 0.75 & \textbf{72.34 $\pm$ 0.79} \\
    32 & 0.0043 $\pm$ 0.0002 & \textbf{0.0031 $\pm$ 0.0002} & 0.002 & 1.000 & 62.34 $\pm$ 1.22 & \textbf{70.18 $\pm$ 2.74} \\
    64 & 0.0049 $\pm$ 0.0003 & \textbf{0.0034 $\pm$ 0.0004} & 0.002 & 1.000 & 58.67 $\pm$ 0.95 & \textbf{65.84 $\pm$ 4.29} \\
    \bottomrule
  \end{tabular*}
\end{table*}

\subsection{Main Accuracy Trends}
Table~\ref{tab:acc_main} summarizes accuracy across seeds ($mean \pm std$). Across all batches (4--64), QFWP consistently outperforms QLSTM in both RMSE and DA. Per batch Wilcoxon tests confirm significance for both metrics (all $p\le 0.004$), with large effect sizes (for RMSE, Cliff's $\delta\ge 0.80$). QLSTM's RMSE increases and DA decreases as batch grows, whereas QFWP remains stable through batch 16 and degrades more slowly thereafter. For example, as the batch size increases from 4 to 64, the mean RMSE for QLSTM rises from $0.00337 \to 0.00487$, while QFWP increases from $0.00298 \to 0.00339$. Over the same range, the mean DA drops from $69.04\% \to 58.67\%$ for QLSTM and $72.72\% \to 65.84\%$ for QFWP. Notably, QFWP maintains both lower error and higher directional skill at every batch, indicating that its advantage is not confined to a single operating point but persists under reduced update frequency. Variance also remains controlled for both models, although DA variance widens for QFWP at batch 64, suggesting that very large batches can increase sensitivity to initialization even when mean performance remains strong.

% ------------------------------------------------
\subsection{Speed and Scaling}
Tables~\ref{tab:time_decomp} and \ref{tab:speedup_decomp} report per component times and speedups (median [IQR]) across batch sizes. Train forward speedups are strong for both models (about 2.15 to 2.42$\times$ across batches). Backward speedup is modest, QLSTM 1.01 to 1.05$\times$ and QFWP 1.02 to 1.22$\times$, and thus caps the end to end full training speedup at about $2\times$. Concretely, the full training median speedup is 2.08$\times$ (QLSTM) vs. 1.94$\times$ (QFWP) at batch 32, and 2.23$\times$ vs. 2.02$\times$ at batch 64. We do not observe any backward slowdowns (below 1$\times$); backward remains the dominant bottleneck. Inference forward speedups stay near 2$\times$ for both models, which is practically relevant because it indicates that batching benefits are not limited to training and can translate to evaluation or deployment throughput when batched queries are available. The IQR summaries further show that timing variability across seeds is small compared to the median gaps, supporting the stability of the reported scaling trends.

% ------------------------------------------------
\subsection{Pareto Tradeoff}
Figure~\ref{fig:pareto_frontier} shows full training speedup (median) against test RMSE (mean), with each point annotated by batch size and model distinct marker styles to improve readability in grayscale. The union of these labeled points traces a clear Pareto frontier:
(1) QFWP contributes the accuracy optimal frontier points up to about 2.0$\times$ speedup (batch 8 to 64); (2) QLSTM extends the throughput end (rightmost points at batch 32 and 64) with higher RMSE.
Therefore, users can pick batch sizes from 8 to 16 for a balanced regime, achieving approximately 1.6 to 1.9$\times$ speedup with minimal RMSE drift for QFWP. Alternatively, batch size 32 may be preferred when prioritizing speed (QLSTM) or error (QFWP). This frontier view emphasizes that batching is a tunable system parameter: increasing batch size moves along a predictable efficiency axis, while the two architectures occupy distinct regions that favor either accuracy (QFWP) or maximum throughput (QLSTM).

% --- (B) Batch-time decomposition (seconds): Train-Forward / Backward / Full-Train / Infer-Forward ---
% New table: 更新時間 2025-10-15 11:00
\begin{table*}[!htbp]
  \centering
  \small
  \caption{Batch implementation wall-clock time (seconds) per batch size: median [Q1, Q3] across seeds.}
  \label{tab:time_decomp}
  \begin{tabular*}{\textwidth}{@{\extracolsep{\fill}}lrrrrr}
    \toprule
    Model & Batch & Train-Forward & Backward & Full-Train & Infer-Forward \\
    \midrule
    QLSTM & 4 & 18.77 [17.71, 18.94] & 1.36 [1.35, 1.38] & 228.51 [226.83, 237.84] & 15.80 [15.10, 16.65] \\
    QLSTM & 8 & 17.67 [16.79, 18.04] & 1.35 [1.34, 1.38] & 195.40 [194.03, 197.88] & 15.38 [14.86, 16.36] \\
    QLSTM & 16 & 17.95 [16.36, 18.13] & 1.33 [1.32, 1.38] & 174.67 [174.10, 176.15] & 16.18 [15.32, 16.57] \\
    QLSTM & 32 & 17.92 [16.49, 20.00] & 1.36 [1.30, 1.38] & 162.63 [160.49, 165.05] & 16.09 [15.05, 16.41] \\
    QLSTM & 64 & 17.61 [16.84, 18.01] & 1.35 [1.31, 1.36] & 153.88 [152.64, 155.67] & 15.92 [15.33, 16.46] \\
    QFWP & 4 & 2.66 [2.63, 2.78] & 0.67 [0.67, 0.68] & 42.23 [41.58, 43.20] & 1.49 [1.47, 1.54] \\
    QFWP & 8 & 2.51 [2.48, 2.59] & 0.79 [0.79, 0.80] & 34.72 [34.41, 35.25] & 1.56 [1.53, 1.58] \\
    QFWP & 16 & 2.54 [2.48, 2.58] & 0.66 [0.66, 0.68] & 31.53 [31.45, 31.63] & 1.48 [1.46, 1.54] \\
    QFWP & 32 & 2.51 [2.48, 2.54] & 0.66 [0.66, 0.67] & 29.23 [28.77, 29.62] & 1.67 [1.65, 1.69] \\
    QFWP & 64 & 2.63 [2.59, 2.67] & 0.67 [0.66, 0.67] & 27.95 [27.54, 28.50] & 1.58 [1.53, 1.61] \\
    \bottomrule
  \end{tabular*}
  \vspace{1mm}

  \footnotesize
  Note: Train-Forward and Backward timings include Autograd graph construction, whereas Infer-Forward (model.eval() with torch.no\_grad()) does not. Full-Train is the total time over two epochs, including all mini-batches.
\end{table*}

% --- (C) Speedup (non-batch/batch): Train-Forward / Backward / Full-Train / Infer-Forward (median [Q1, Q3]) ---
% New table: 更新時間 2025-10-15 11:00
\begin{table*}[!t]
  \centering
  \small
  \caption{Speedup (non-batch / batch) across batch sizes: median [Q1, Q3] over seeds.}
  \label{tab:speedup_decomp}
  \begin{tabular*}{\textwidth}{@{\extracolsep{\fill}}lrrrrr}
    \toprule
    Model & Batch & Train-Forward & Backward & Full-Train & Infer-Forward \\
    \midrule
    QLSTM & 4 & 2.15 [2.00, 2.20] & 1.01 [0.98, 1.03] & 1.41 [1.41, 1.43] & 2.05 [1.88, 2.19] \\
    QLSTM & 8 & 2.20 [2.12, 2.31] & 1.03 [0.98, 1.05] & 1.67 [1.66, 1.69] & 2.02 [1.94, 2.35] \\
    QLSTM & 16 & 2.29 [2.06, 2.45] & 1.05 [1.01, 1.07] & 1.90 [1.89, 1.92] & 1.98 [1.89, 2.17] \\
    QLSTM & 32 & 2.20 [2.02, 2.36] & 1.03 [1.01, 1.05] & 2.08 [2.03, 2.13] & 2.17 [1.95, 2.24] \\
    QLSTM & 64 & 2.42 [2.25, 2.57] & 1.05 [1.03, 1.07] & 2.23 [2.20, 2.48] & 2.10 [2.07, 2.16] \\
    QFWP & 4 & 2.27 [2.11, 2.39] & 1.19 [1.18, 1.20] & 1.37 [1.34, 1.39] & 2.32 [2.19, 2.39] \\
    QFWP & 8 & 2.30 [2.21, 2.40] & 1.02 [1.02, 1.03] & 1.63 [1.61, 1.64] & 2.22 [2.18, 2.40] \\
    QFWP & 16 & 2.34 [2.22, 2.41] & 1.20 [1.19, 1.22] & 1.77 [1.74, 1.79] & 2.29 [2.12, 2.39] \\
    QFWP & 32 & 2.35 [2.26, 2.39] & 1.22 [1.19, 1.22] & 1.94 [1.87, 2.00] & 2.08 [2.01, 2.15] \\
    QFWP & 64 & 2.27 [2.23, 2.31] & 1.20 [1.18, 1.21] & 2.02 [1.95, 2.06] & 2.18 [2.15, 2.26] \\
    \bottomrule
  \end{tabular*}
\end{table*}

\section{Discussion}
\label{sec:discussion}

\textbf{Forward Backward Asymmetry and Its Systemic Implications.}
Across all experiments we observe a pronounced asymmetry between forward and backward scaling under adjoint differentiation. With PennyLane state vector simulation and analytical reverse mode gradients, the forward path benefits strongly from batching because evaluating multiple samples in a single QNode call with parameter broadcasting reduces Python and QNode overhead and improves memory locality, resembling kernel fusion in classical deep learning. In contrast, the backward pass is dominated by the adjoint reverse time sweep that reconstructs and traverses the full computational graph. Its cost scales mainly with circuit depth, trainable parameter count, and unrolled sequence length, and only weakly with batch size. We therefore expect the asymmetry to increase for longer sequences or deeper circuits, and leave controlled sweeps over sequence length and quantum depth for future work. Empirically, forward speedup often exceeds $2\times$ (about 2.2 to 2.4$\times$), whereas backward scaling saturates near 1.0 to 1.22$\times$ (QLSTM 1.01 to 1.05$\times$, QFWP 1.18 to 1.22$\times$). The resulting end to end training speedup of about $2\times$ matches an Amdahl style upper bound where the backward fraction limits overall gain~\cite{amdahl1967validity}. This indicates that gradient computation remains the main performance limiter. Potential remedies include alternative differentiators such as parameter shift with batched shifted circuits, or stochastic estimators on hardware, as well as update frequency control when large batches reduce step counts. QFWP shows slightly better backward scaling despite similar circuit depth, consistent with a smaller differentiation graph because its trainable parameters are classical and the quantum layer is driven by a compact slow programmer.

\textbf{Accuracy Dynamics and Optimization Robustness.}
Batch size also changes accuracy through reduced update frequency under our fixed two epoch budget without gradient accumulation. QLSTM degrades rapidly as batch size increases from 4 to 64, with RMSE rising from $0.00337$ to $0.00487$ and DA dropping from $69.0\%$ to $58.7\%$. QFWP remains stable through batch 16 (RMSE $\approx0.0029$, DA $\approx72.3\%$) and declines more slowly thereafter (batch 64 RMSE $\approx0.0034$, DA $\approx65.8\%$). Across batches, paired Wilcoxon signed rank tests yield $p\le0.004$ with large effect sizes (Cliff's $\delta\ge0.80$), confirming a robust QFWP advantage. We attribute this to architectural decoupling: the slow programmer provides temporal coherence, and the outer product accumulation of fast parameters acts as an implicit stabilizer when updates are less frequent. More broadly, hybrid designs that delegate temporal structure to classical controllers appear more robust under throughput constrained training, while fully quantum gated recurrence is more sensitive to reduced backpropagated update counts.

\textbf{Pareto Frontier and Design Guidelines for Practitioners.}
Figure~\ref{fig:pareto_frontier} summarizes the trade off between accuracy and throughput. Over all batch sizes, QFWP occupies the accuracy optimal region (RMSE $<0.0032$, DA $>70\%$) for speedups below $2\times$, while QLSTM dominates the highest throughput region above $2\times$ with higher error. This suggests complementary operating regimes rather than direct competition. For accuracy critical or inference sensitive workloads, QFWP with batch 8 to 16 offers the best balance (RMSE $\approx2.9\times10^{-3}$, DA $\approx72\%$, full training speedup $\approx1.7\times$, inference forward $\approx2.2\times$). For throughput oriented scenarios, QLSTM with batch 64 achieves maximum training throughput ($2.23\times$ [2.20, 2.48]) with reduced accuracy. For mixed workloads, QFWP with batch 32 provides a practical middle point (RMSE $\approx3.1\times10^{-3}$, DA $\approx70.2\%$) with about $1.9\times$ speedup. These heuristics treat batching as a system level knob that jointly controls efficiency, gradient cost, and accuracy.

%  ===============================================
\section{Conclusion}
\label{sec:conclusion}
This work presents a numerically grounded and EPC aligned comparison between QLSTM and QFWP under non batched and batched training, using adjoint differentiation on real EUR/USD forecasting data. Batching accelerates the forward pass but offers limited backward scaling, so full training speedup saturates near $2\times$ by batch 32. Under a fixed two epoch budget, QFWP achieves lower RMSE and higher DA across all batch sizes (Wilcoxon $p\le0.004$, Cliff's $\delta\ge0.80$), while QLSTM attains the highest throughput at batch 64. Together, they form a speed accuracy Pareto frontier, with QFWP favoring accuracy and QLSTM favoring throughput. Beyond head to head results, this study provides a system level benchmark that isolates batch effects under adjoint differentiation and clarifies how quantum gated recurrence and hybrid controllers behave when throughput, precision, and gradient cost must be co optimized. The results suggest that scaling will depend on improved differentiators and training mechanics, not circuit design alone, and that EPC alignment is a practical diagnostic for deployable efficiency. Limitations include a single 1D financial dataset and simulator only backends. Future work will extend EPC benchmarking to other modalities and architectures.

\section*{Acknowledgements}
The authors conducted all experiments on simulator backends that reflect state-vector dynamics rather than hardware noise, so timings represent relative efficiency. This work used PennyLane \texttt{lightning.qubit} (adjoint backend) on Windows~11 + WSL2 (Ubuntu~24.04.1~LTS) with an Intel~Core~Ultra~7~265K and 96~GB~RAM.

The views expressed are those of the authors and do not represent Wells Fargo. This article is for informational purposes only and should not be construed as investment advice. Wells Fargo makes no express or implied warranties and disclaims any legal, tax, or accounting implications related to this article.

\bibliographystyle{IEEEbib}
\bibliography{strings,refs}

\end{document}